\documentstyle[preprint,pra,aps]{revtex}
\begin{document}
\draft
\title{First order transition from ferro- to                
antiferromagnetism in CeFe$_2$ based pseudobinary alloys}
\author{Meghmalhar Manekar, S. B. Roy and P. Chaddah}
\address{Low Temperature Physics Laboratory,\\
Centre for Advanced Technology,\\ Indore 452013, India}
\date{\today}
\maketitle
\begin{abstract}
We present results of ac susceptibility measurements highlighting the 
presence of thermal hysteresis and phase coexistence 
across the ferro-to antiferromagnetic transition in various CeFe$_2$ based
pseudobinary systems. These results indicate that the ferro-to antiferromagnetic
transition in these systems is first order in nature.
\end{abstract}                          
The C15-Laves phase compound CeFe$_2$ retains  its 
identity amongst the other members of RFe$_2$ family (where 
R=Y,Zr 
and heavy rare earth elements). First, magnetic moment of   
CeFe$_2$ per formula unit ($\approx 2.4 \mu_B$) is distinctly   
smaller than that found in other RFe$_2$ compounds\cite{1}.
Second,its Curie temperature T$_C$($\approx$ 235 K) 
is relatively small in 
comparison to the T$_C$ of other RFe$_2$ compounds 
\cite{1}.However, short range magnetic order is detected in   
its paramagnetic state even in the temperature regime upto 
four times T$_C$ \cite{2}.All these aspects drew 
the attention of the experimentalists during last     
thirty years, and amongst other things 
the role of Ce 
in the magnetic properties of CeFe$_2$ has been a subject of both
theoretical\cite{3} and experimental investigations \cite{4,5};this in 
turn led to the discovery of newer interesting properties \cite{6}. 
The most recent neutron measurements on 
single crystal sample of pure CeFe$_2$ have now revealed the 
presence of low temperature antiferromagnetic fluctuation    
in this otherwise ferromagnetic compound \cite{6}. From the   
study of doped-CeFe$_2$ it is already known for quite 
sometime that the ferromagnetism of CeFe$_2$ is quite fragile in     
nature and a low temperature antiferromagnetic state can be 
established easily with small amount of doping with elements 
like Al,Co,Ru,Ir,Re,Os \cite{7,8,9,10,11,12,13}. 
It should be noted, however, 
that the destabilization of ferromagnetism in CeFe$_2$ is 
not a simple disorder induced one, since the doping with 
other elements like Ni,Mn,Rh,Pd leads to simple dilution of 
ferromagnetism \cite{8,10}.

Most of the early experimental activities in CeFe$_2$ were focussed to 
establish the exact nature of the low temperature magnetic 
phase, whether it is a re-entrant spin-glass \cite{14,15} or an 
antiferromagnetic state \cite{9,10,16,17} and except 
in few cases \cite{17,18} 
not much emphasis was given on the exact nature of 
this phase transtion. With the antiferromagnetic nature of the low 
temperature state being more or less established \cite{16,17}, 
in the present work we shall  
specifically address the question --what is  the nature of this 
ferro- to antiferromagnetic transition? 
While there exists no complete theory (to our knowledge) to 
explain the interesting magnetic properties of CeFe$_2$, a 
phenomenological model  dealing with itinerant electron 
systems \cite{19} has often been invoked to explain the 
para-to ferro- to antiferromagnetic transition in the doped-psuedobinary 
alloys of CeFe$_2$.This phenomenological model of   
Moriya and Usami predictd 
that the ferro- to antiferromagnetic transition 
would be a first order transition, while para- to 
ferromagnetic transition would be a second order transition \cite{19}. 
With our high       
resolution ac-susceptibility measurement across this 
ferro- to antiferromagnetic transition in two doped 
samples of CeFe$_2$, we shall report characteristics 
which are typically associated with a first order transition.On     
the other hand the higher temperature para- to 
ferromagnetic transition can be characterized as a standard      
second order phase transtion. We believe that such a clear cut   
characterization of the various phase transitions in          
CeFe$_2$ based pseudobinaries is necessary, either for an     
appropraite extension of Moriya-Usami model \cite{19} or for the
development of newer theory for the proper understanding   
of the magnetic properties of CeFe$_2$.

Two samples--Ce(Fe,5\%Ir)$_2$ and Ce(Fe,7\%Ru)$_2$--used in the present study were 
prepared by argon arc melting from metals of at
least 99.99\% purity.Details of sample preparation, heat treatment and characterisation
can be  found in Ref. 10.The same samples have earlier 
been used in some other studies \cite{13,20,21}.

The AC susceptibility setup consists of a coil system having a primary
solenoid and two oppositely wound secondaries each consisting of 1500 turns.
The coil is dipped in liquid nitrogen to 
ensure that the temperature of the coil remains constant during
the entire experiment to avoid drifts in the value of the applied field.
The sample is mounted in a double walled
quartz insert and its
temperature is raised by heating the exchange gas by a
heater wound on a seperate teflon mounting.A temperature controller 
(Lake Shore-- DRC-91CA)
is used for controlling the temperature.A copper-constantan
thermocouple is used in differential mode to monitor the 
small temperature lag between the sample and the sensor.The
sinusoidal output of a lock in amplifier 
(Stanford Research--SR830) is given to a voltage
to current convertor which drives the current through the 
coil to generate the neccessary ac magnetic field.
The signal from the pickup coil
which is proportional to the susceptibility is measured 
by the same lock in amplifier.  
The field and frequency values were 4 Oe rms and 333Hz respectively.

Fig.(1) shows the AC susceptibility ($\chi$) for both Ce(Fe,5\% Ir)$_2$ 
and Ce(Fe,7\% Ru)$_2$ 
as a function of temperature (T).
The para- to ferromagnetic transition is characterized by a sharp 
increase in susceptibility ($\chi$) with the decrease in T 
at  T$_{Curie}\approx $185K 
in the 5\% Ir doped sample and T$_{Curie}\approx$165K in the 7\% Ru doped sample.
Below T$_{Curie}$ the susceptibility more or less flattens out for both
the samples, before decreasing sharply 
at around 135K in Ce(Fe,5\% Ir)$_2$ and at around 125K in Ce(Fe,7\% Ru)$_2$.
This low temperature decrease in $\chi$ has earlier been taken 
as a signature of ferro- to antiferromagnetic transition \cite{10,13,20}, 
and the estimated transition temperatures (T$_N$) from our present study
agree well with the existing literature \cite{10,13,20}.

Our aim now is to find out the exact nature of these two magnetic 
transitions oberved in CeFe$_2$-based pseudobinaries. 
Experimentally, the indication of a first order transition usually comes
via a hysteretic behaviour of various properties, 
not necessarily thermodynamic ones. As an example, the first indication of a
first order melting transition from elastic solid to vortex liquid
in a vortex matter came via distinct hysteresis observed in transport 
property measurements \cite{22,23}.  
The confirmatory tests of the first order nature of a transition 
ofcourse involve the detection of discontinuous change in thermodynamic
observables and the estimation of latent heat, and this has subsequently
been achieved for vortex lattice melting in vortex matter \cite{24,25}. 
There also exists a less rigorous class of experimental tests which 
invloves the study of phase inhomogeneity and phase coexistence 
across a first order transtion. This kind of experiment has also come
out to be pretty informative for the melting transtion \cite{26} as well as
ordered solid to disordered solid transition \cite{27,28} 
in vortex matter. In our present study we shall use hysteresis and phase coexistence
to investigate the nature 
of the magnetic transitions in CeFe$_2$ based systems; our observable will be
ac susceptibility ($\chi$).

In order to observe a hysteresis in the transition, if any, 
we have chosen to sweep the temperature
at a slow rate (0.006K/sec typical and slower when needed) 
instead of stabilizing at each temperature.
This was done to ensure that the temperature is varied unidirectionally 
during both the heating and cooling cycles.
The signal was measured at a temperature interval of 0.2K.
The time constant of the low pass filter
of the lock in amplifier was chosen such that the temperature 
changes negligibly(compared to our temperature step)
within a time interval of 10 times the time constant.
The temperature difference between the sensor and the sample,
as monitored by the differential thermocouple, was always less than 
1\% of the sensor temperature and is used to obtain the correct value of 
the sample temperature.

First, we show the effect of temperature cycling on the para- 
to ferromagnetic transition in fig(2).In case of 
Ce(Fe,5\% Ir)$_2$ the transition is reversible within an error 
of 0.15K to 0.2K.In case of Ce(Fe,7\% Ru)$_2$ the reversibility is even better.
The lack of hysteresis in para- to ferromagnetic transition 
within an error bar smaller than our temperature step,
is indicative of a second order phase transition.

We then focus our attention on the ferro- to antiferromagnetic transition 
which has been shown to be associated
with a structural distortion from cubic to rhombohedral \cite{16,17}, 
hinting towards a first order transition.The same protocol
of sweeping the temperature and measuring the signal at 
closely spaced temperature values is followed during this measurement also.

Fig(3) shows the result of our measurements on both 5\% Ir and 7\% Ru 
doped CeFe$_2$ samples. Both the samples show
a distinct thermal hysteresis in the ac-susceptibility across the 
ferro- to antiferromagnetic transition.The width of the hysteresis is about 2K 
which is well beyond the error in our measurements.

To study the phase coexistence we use the technique of minor hysteresis
loop (MHL)\cite{29}. We first define the ``envelope curve" as the curve enclosing 
the thermally hysteretic susceptibility 
beteween the lower and higher temperature  
reversible region (see Fig.3).
We can draw a MHL during the heating cycle 
i.e. start heating and increase T from the lower temperature reversible 
(antiferromagnetic) region and then reverse the direction of 
temperature before reaching the higher temperature reversible
(ferromagnetic) region. We can also draw a MHL in the 
cooling cycle i.e. start cooling from the 
reversible ferromagnetic region and reverse the direction of 
temperature before reaching the lower temperature reversible
antiferromagnetic region.  
If the heating is reversed at sufficiently `low' temperatures 
the minor loop does not coincide with the cooling part of the 
`envelope curve'. Here in the lower part of the hysteretic regime the high
temperature ferromagnetic phase phase is not  formed in a sufficient quantity; so 
when the temperature is decreased the curve does not fall 
on the cooling part of the envelope curve which represents the 
curve along which the high temperature phase is supercooled. 
The MHL's initiated from temperatures well inside the hysteretic regime 
coincide with the cooling 
part of envelope curve indicating that the high temperature phase 
has formed in a sufficient quantity. In Fig. 4 and 5 we present some
representative MHLs both for the Ce(Fe,5\%Ir)$_2$ and Ce(Fe,7\%Ru)$_2$
alloys.
We have drawn similar MHLs from the cooling branch of the enevelope
curve, which are not shown here for the sake of clarity and conciseness. 
We have reproduced this behaviour of MHLs over many experimental cycles.
The presence of these MHLs clearly suggest 
the existence of phase coexistence across the
ferro- to antiferromagnetic transition.
Had there been no phase coexistence we would have followed 
the cooling part of the envelope curve 
reversibly on increasing T.Very similar 
minor hysteresis loop technique has been used to study the 
phase coexistence associated with a first order 
metal-insulator transitions in NdNiO$_3$ \cite{30}.

It should be noted here that the pinning of solitons (domain walls) 
by lattice defects can also give rise to a thermal hysteresis \cite{31} 
in magnetic
measurements. However, the observed thermal hysteresis in our present study
is confined to a relatively narrow temperature window and this argues against
such a possibility.

In conclusion we have shown that the ferro- to              
antiferromagnetic transition in the compounds               
Ce(Fe,5\%Ir)$_2$ and Ce(Fe,7\%Ru)$_2$ is accompanied by     
distinct thermal hysteresis as well as signatures of 
phase-coexistence. We argue that these observations 
are indicative of the first 
order nature of the concerned phase transition. 
The higher tempeature para- to ferromagnetic 
transition appears to be a typical second order phase 
transition. These results would support the 
applicability of Moriya-Usami's model \cite{19} 
in explaining the double magnetic
transitions in various CeFe$_2$ based pseudobinary systems. 
A calorimetric study is now required to confirm the
conjecture that this ferro- to antiferromagnetic transition is 
first order in nature.
However, it should be noted that 
in the case of small latent heats it might be difficult 
to distinguish a first order transition through calorimetric 
studies \cite{32}; in 
such cases the observed hysteresis and phase coexistence 
would remain a useful tool for identification of a first order transition. 

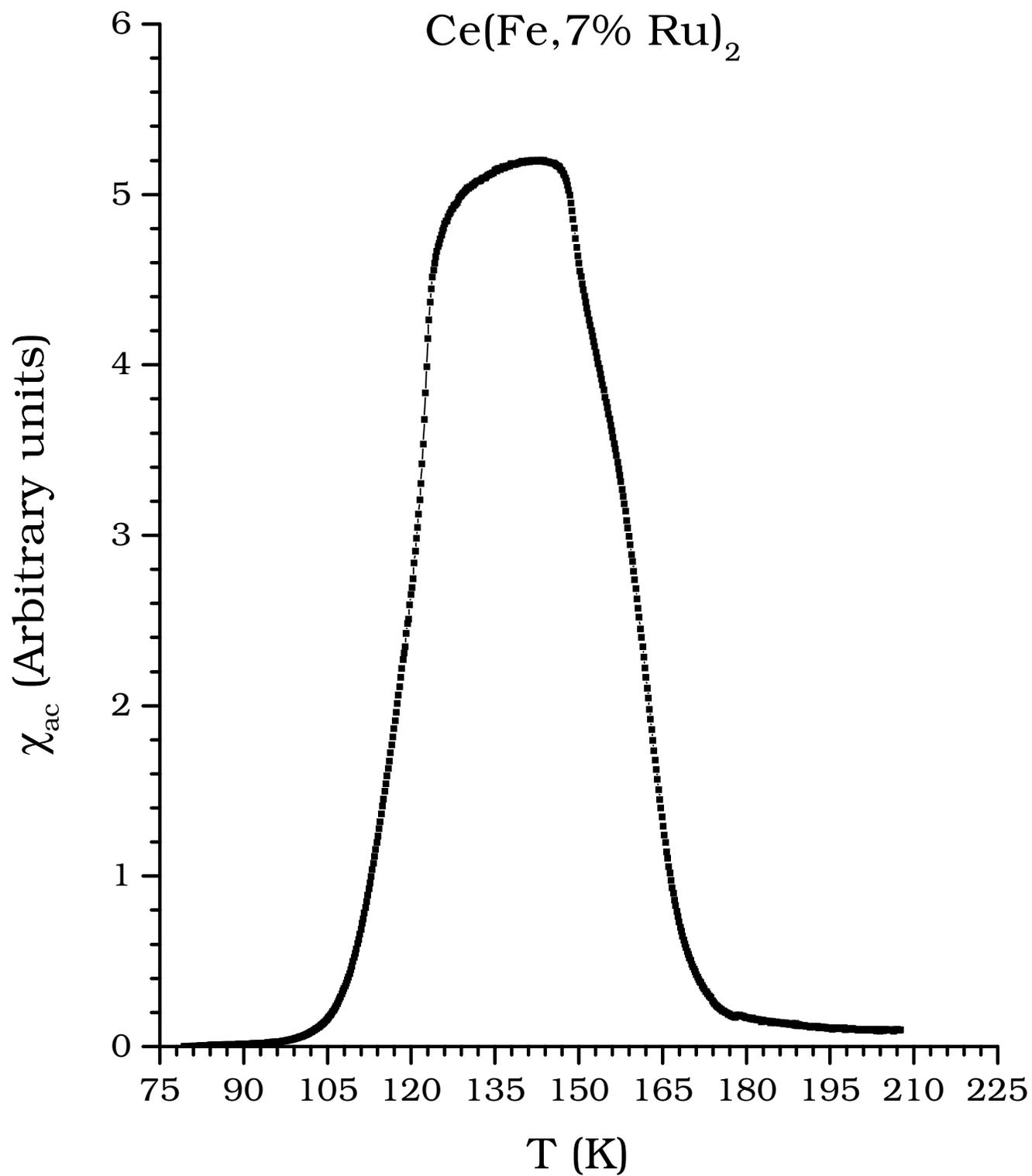
\begin{figure}
\caption{AC susceptibility ($\chi$) versus temperature (T) plots for 
(a)Ce(Fe,5\%Ir)$_2$ (b)Ce(Fe,7\%Ru)$_2$.}
\end{figure}
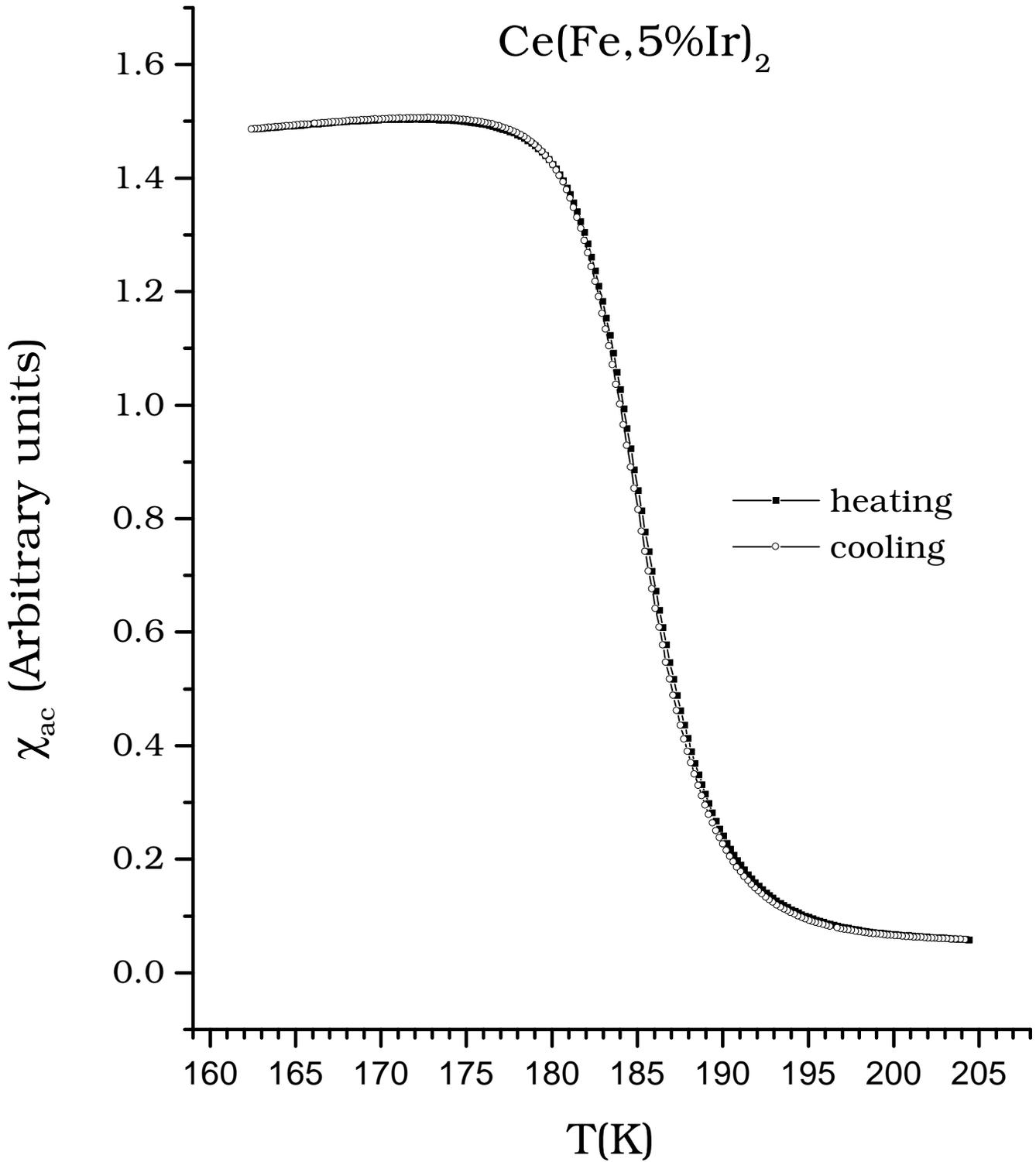
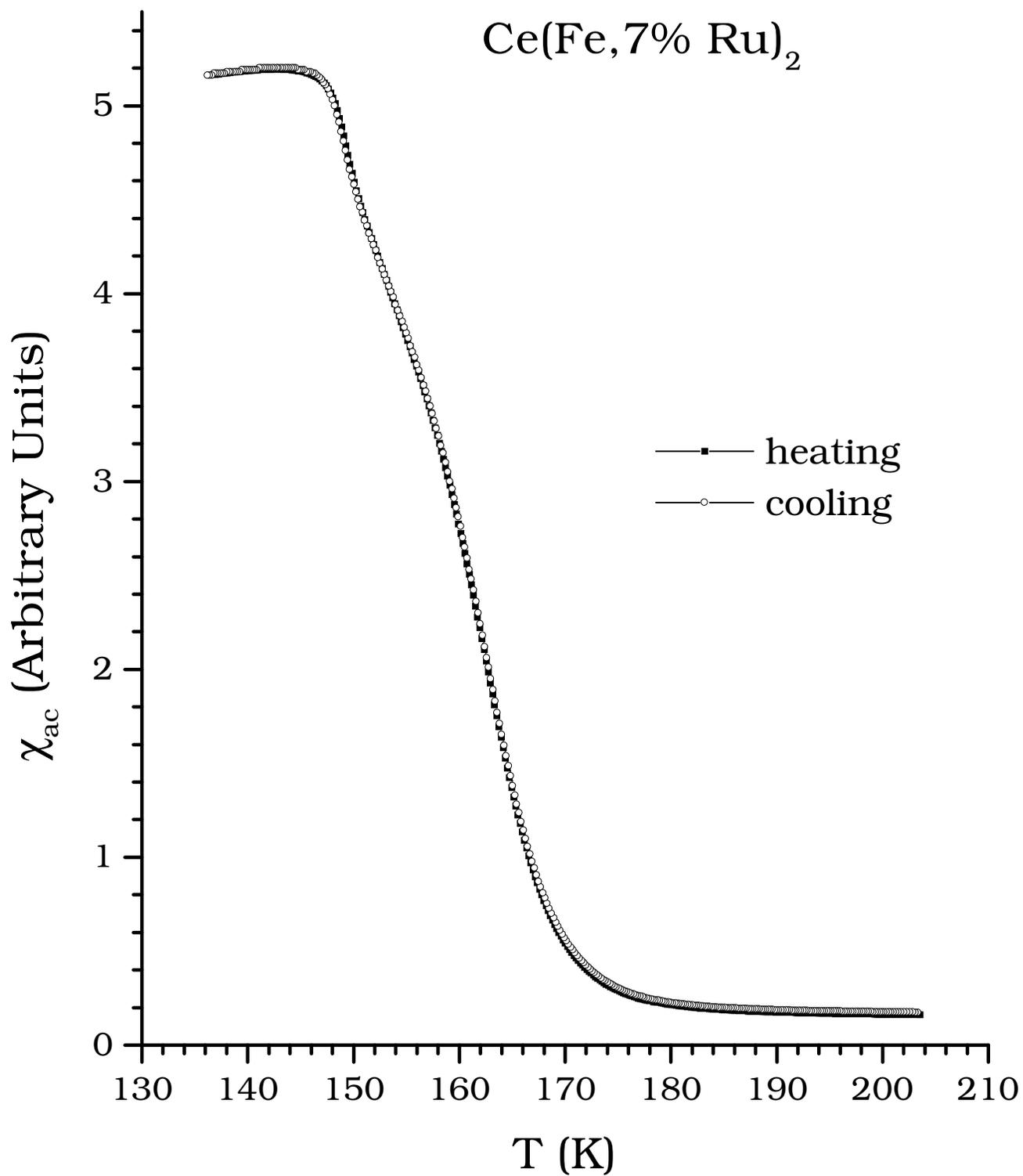
\begin{figure}
\caption{$\chi$ vs T plot highlighting the thermal reversibility 
of the para- to ferromagnetic transition 
in (a)Ce(Fe,5\%Ir)$_2$ (b)Ce(Fe,7\%Ru)$_2$.}
\end{figure}
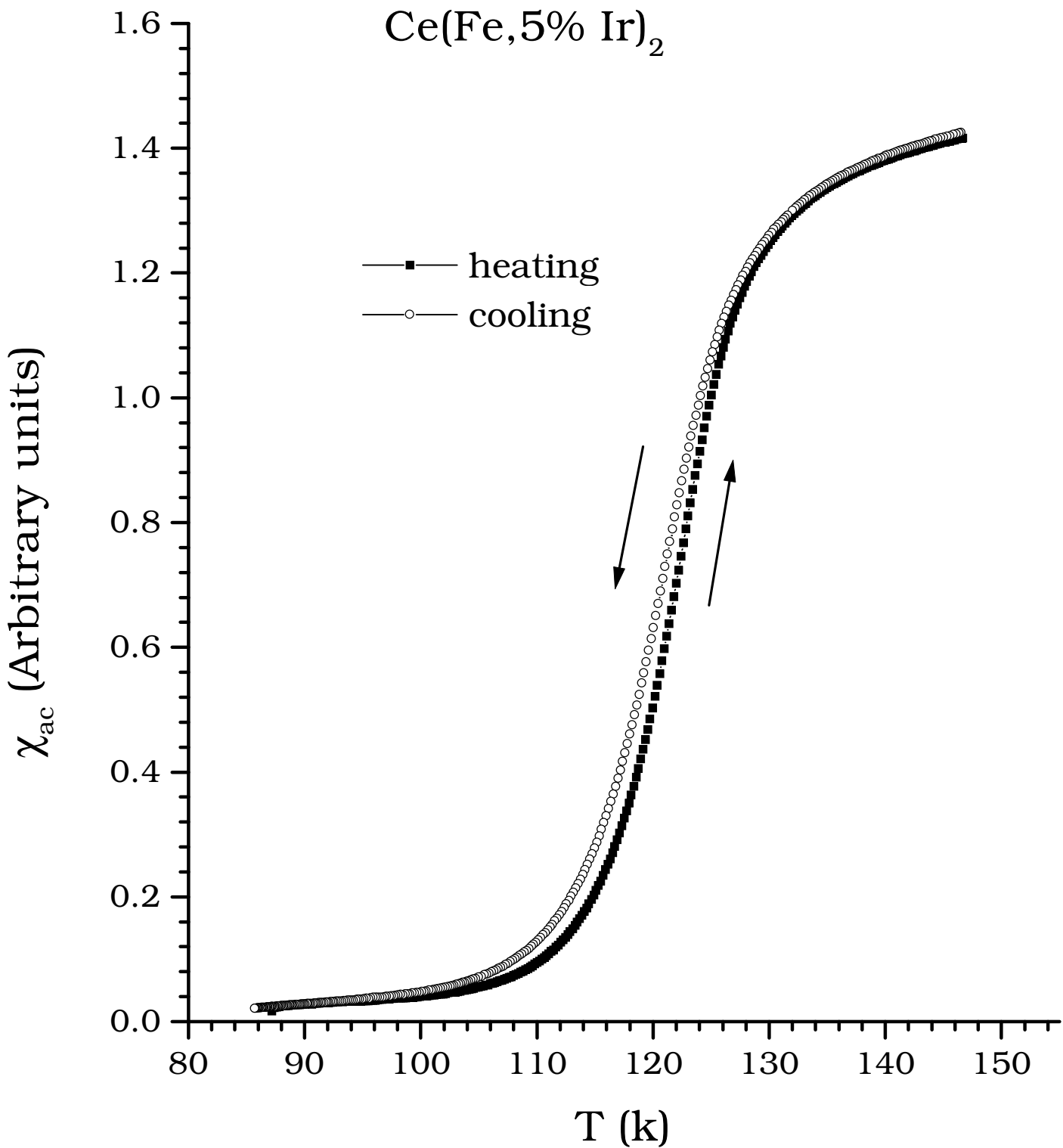
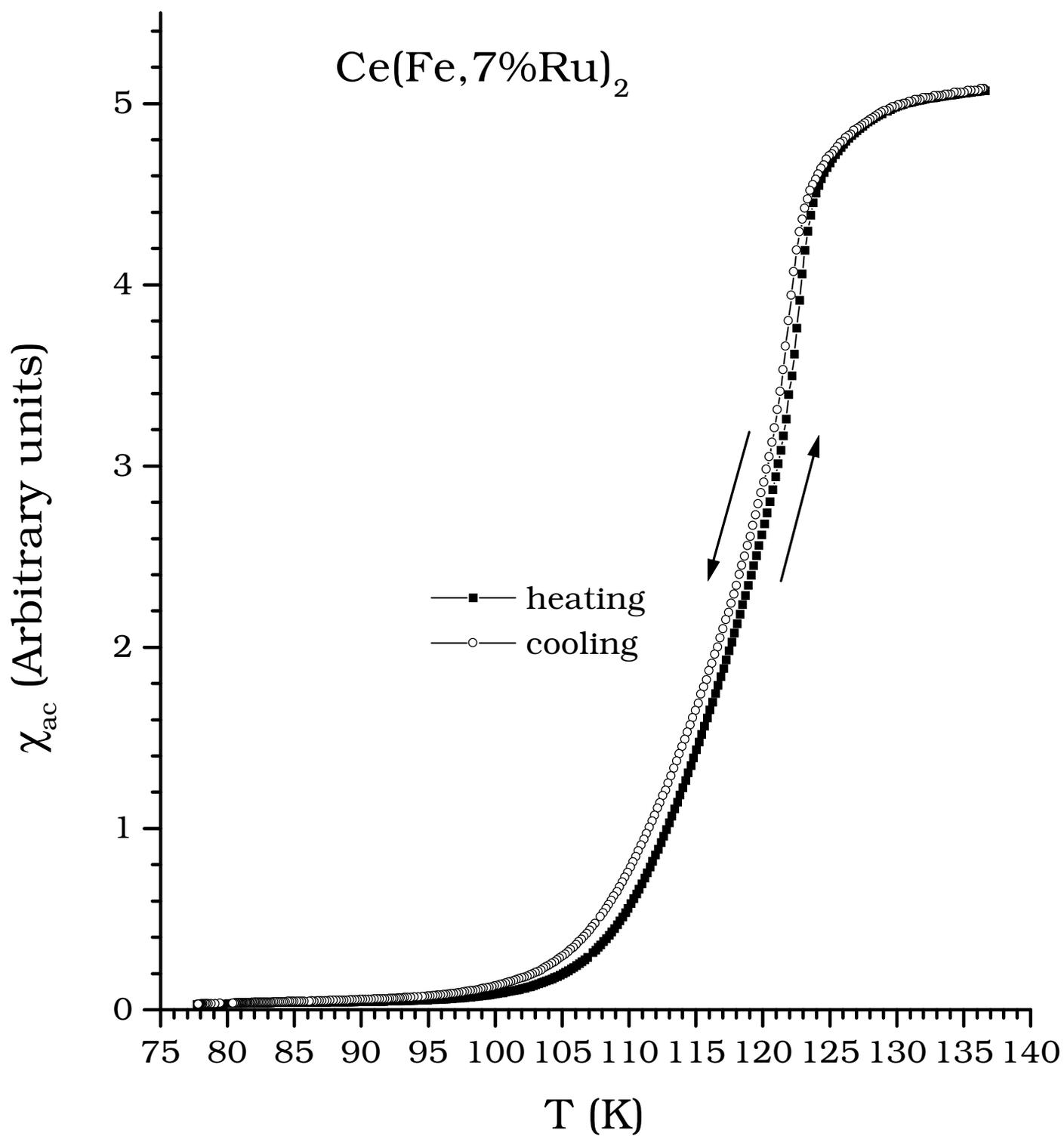
\begin{figure}
\caption{$\chi$ vs T plot highlighting the thermal irreversibility 
of the ferro- to antiferromagnetic transition 
in (a)Ce(Fe,5\%Ir)$_2$ (b)Ce(Fe,7\%Ru)$_2$.}
\end{figure}
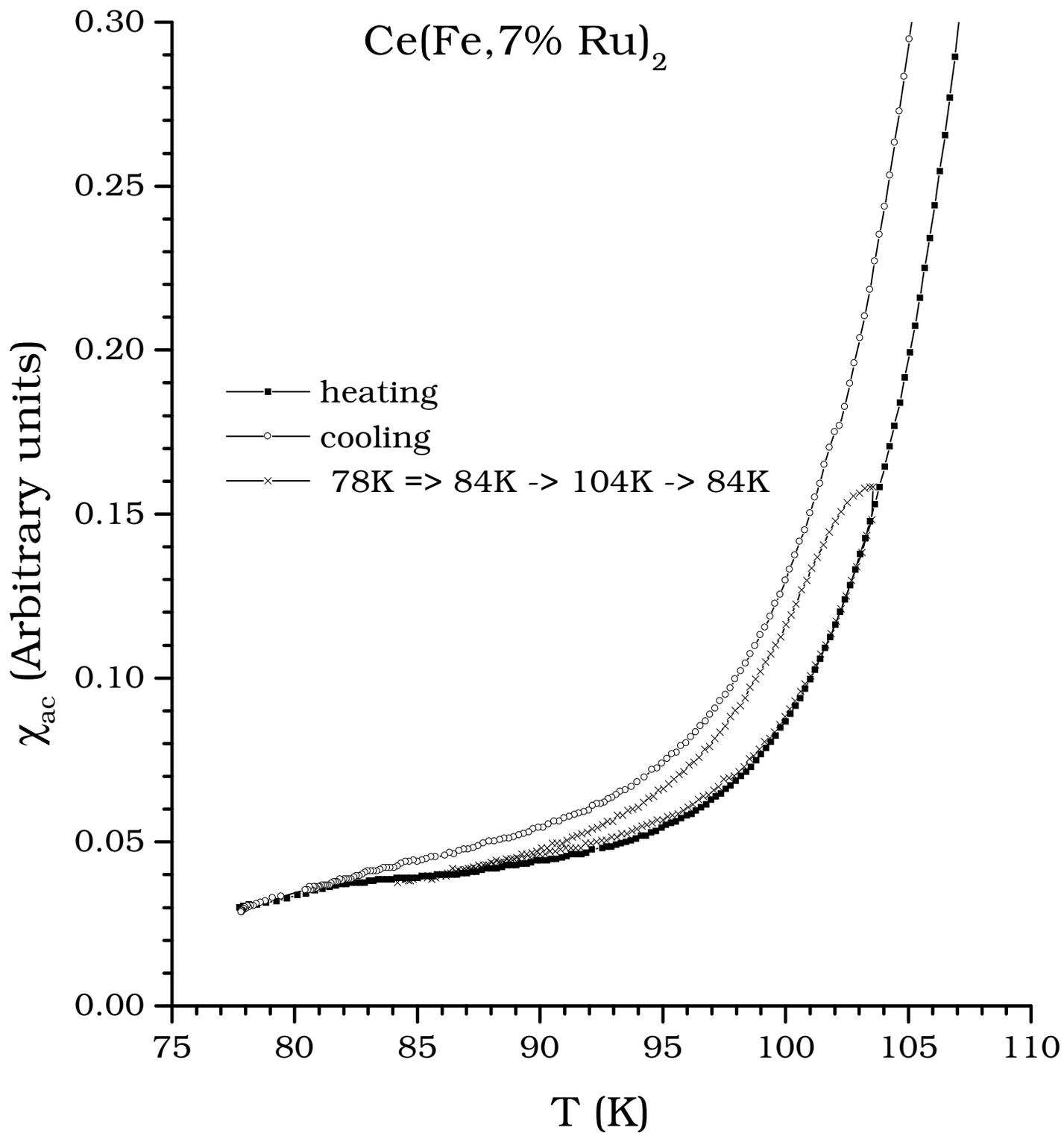
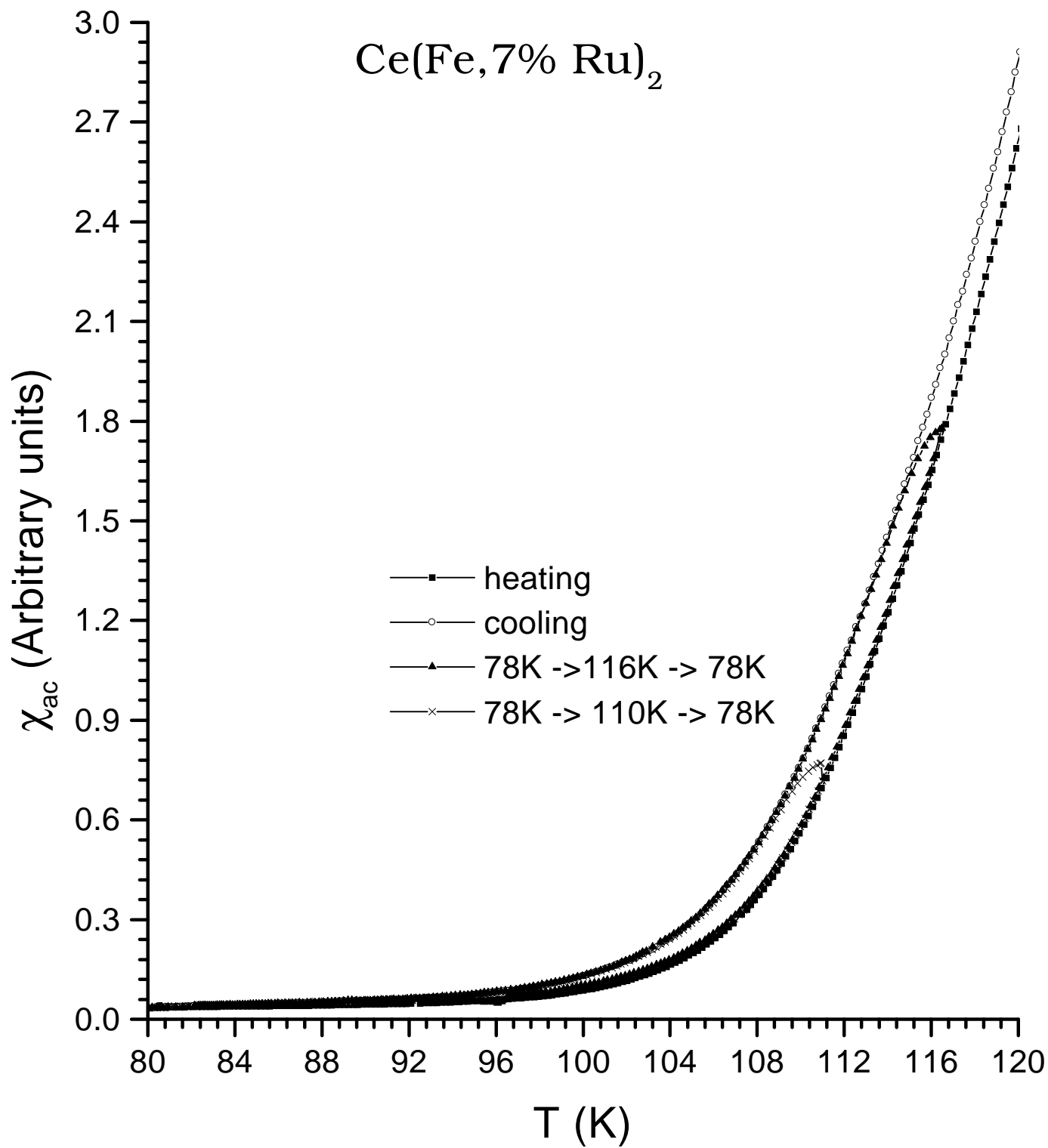
\begin{figure}
\caption{Minor hysteresis loops (MHL) in $\chi$ vs T plot highlighting phase
coexistence in Ce(Fe,7\%Ru)$_2$ : (a)representative MHL initiated from the 
lower part of the hysteretic regime, (b)representative MHLs initiated from  
well inside the hysteretic regime. See text for details.}
\end{figure}
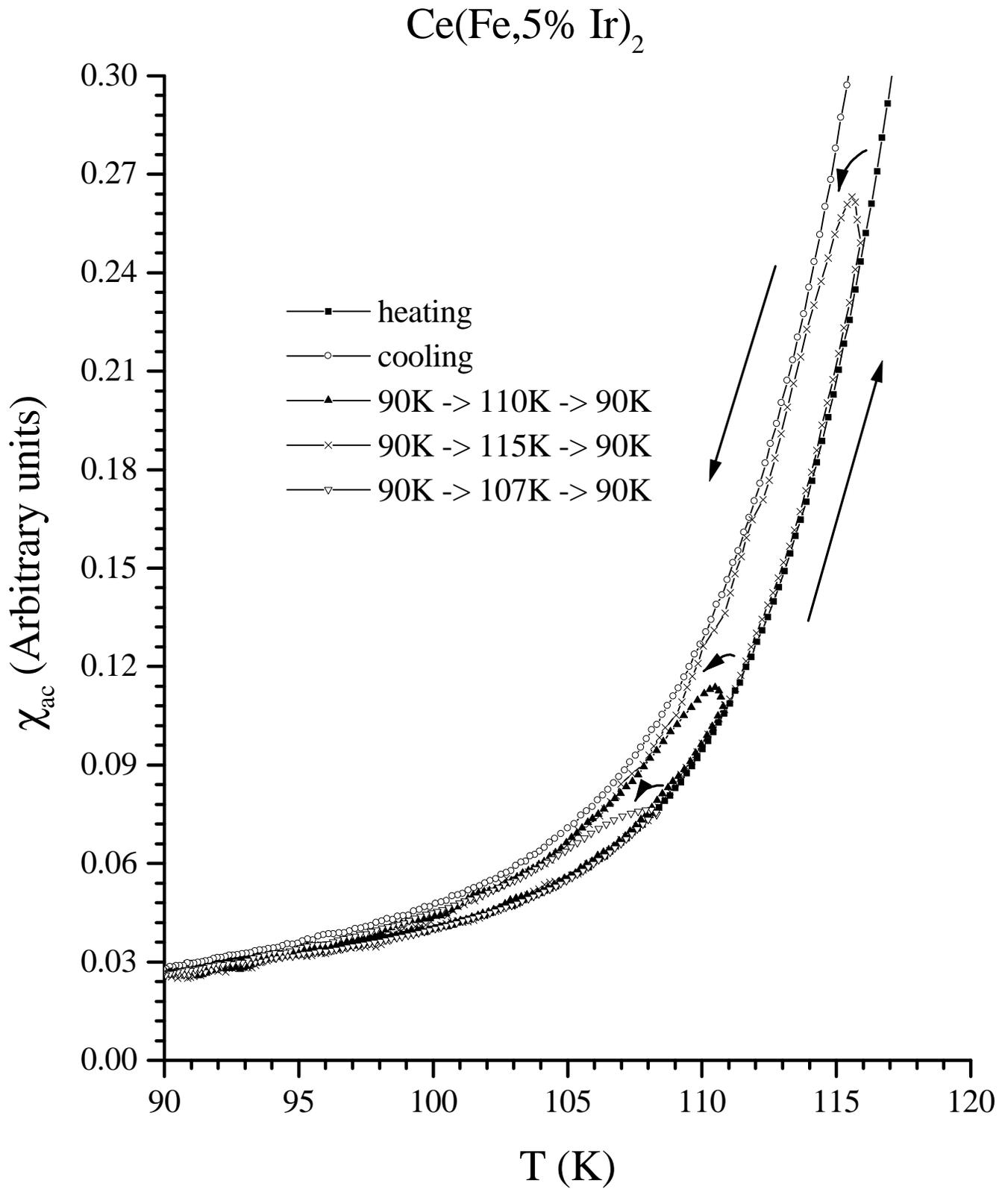
\begin{figure}
\caption{Minor hysteresis loops (MHL) in $\chi$ vs T plot highlighting phase
coexistence in Ce(Fe,5\%Ir)$_2$.}
\end{figure}
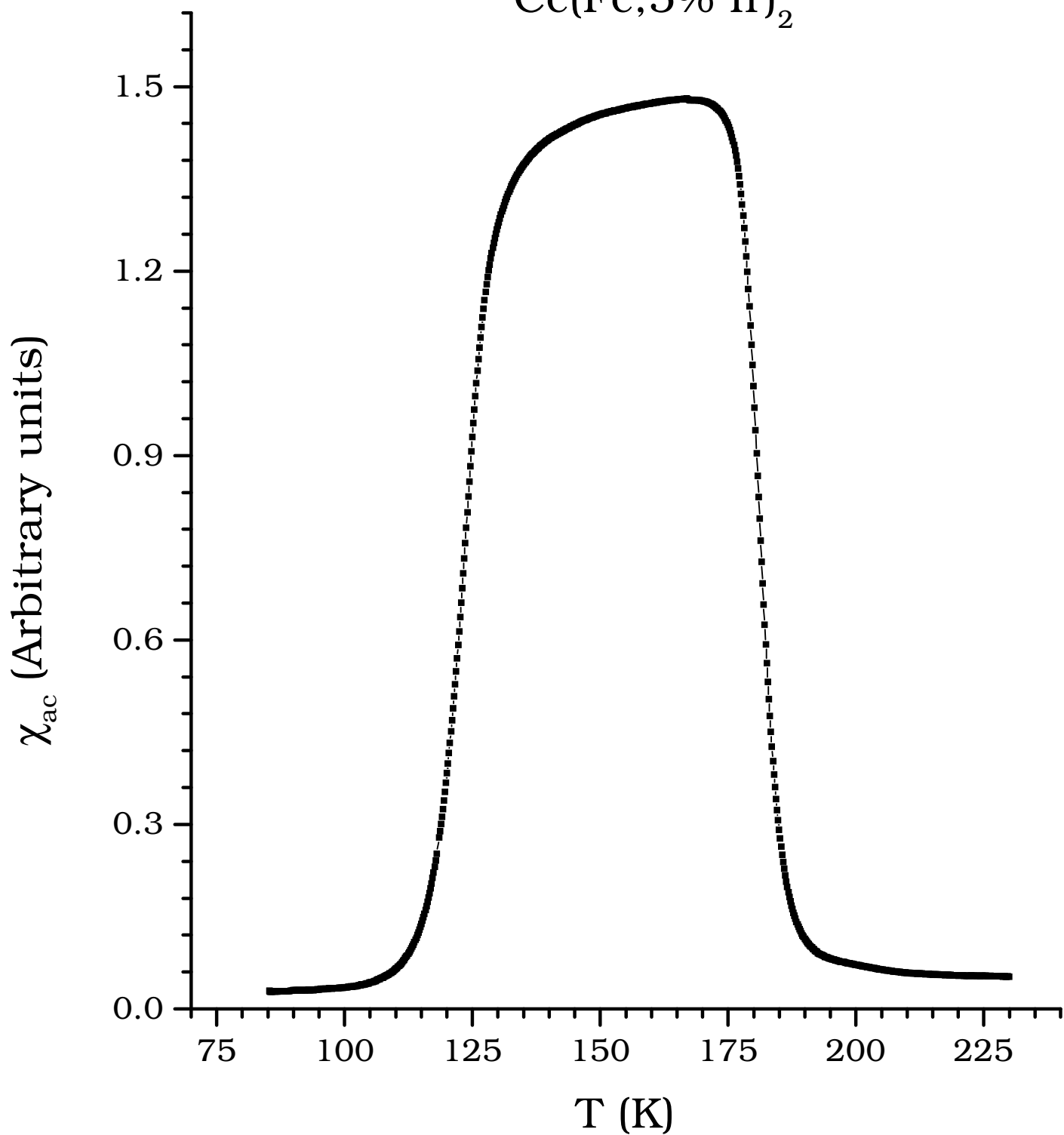
\end{document}